\documentclass[preprint,aps,showpacs,nofootinbib,tightenlines]{revtex4-1}
\usepackage{amsmath}
\usepackage{amssymb}
\usepackage{amsfonts}
\usepackage{CJK}
\usepackage{graphicx,epsfig}
\usepackage{psfrag}
\usepackage{fullpage}
\usepackage{appendix}
\begin{document}
\textheight=230mm
\textwidth=160mm
\topmargin=-0.3in
\unitlength=12mm
\newcommand{\beq}{\begin{equation}}
\newcommand{\eeq}{\end{equation}}
\newcommand{\bea}{\begin{eqnarray}}
\newcommand{\eea}{\end{eqnarray}}
\newcommand{\beal}{\begin{aligned}}
\newcommand{\eeal}{\end{aligned}}
\newcommand{\non}{\nonumber\\ }

\def \cpc{{\bf Chin. Phys. C } }
\def \ctp{{\bf Commun. Theor. Phys. } }
\def \epjc{{\bf Eur. Phys. J. C} }
\def \jhep{ {\bf JHEP } }
\def \jpg{ {\bf J. Phys. G} }
\def \mpla{ {\bf Mod. Phys. Lett. A } }
\def \npb{ {\bf Nucl. Phys. B} }
\def \plb{ {\bf Phys. Lett. B} }
\def \pr{ {\bf Phys. Rep. } }
\def \prd{ {\bf Phys. Rev. D} }
\def \prl{ {\bf Phys. Rev. Lett.}  }
\def \ptp{ {\bf Prog. Theor. Phys. }  }
\def \zpc{ {\bf Z. Phys. C}  }



\title{The loop effects on the chargino decays
$\tilde{\chi}_1^\pm\to \tilde{\chi}_1^0 f f^\prime$  in the MSSM }
\author{Qing-jun Xu$^{1,3}$\footnote{xuqingjun@hznu.edu.cn},
Shu-sheng Xu $^{2}$, Zhen-jun Xiao $^{2}$
\footnote{xiaozhenjun@njnu.edu.cn}  and Li-gang Jin$^{2}$ }
\affiliation{1. Department of Physics,Hangzhou Normal University, Hangzhou
310036, P.R.China\\
2. Department of Physics and Institute of Theoretical Physics,
Nanjing Normal University, Nanjing, Jiangsu 210023, P.R. China\\
3. State Key Laboratory of Theoretical Physics, Institute of Theoretical Physics, Academia Sinica, Beijing 100190, China}
\date{\today}
\begin{abstract}
The lighter chargino three body decays $\tilde{\chi}_1^\pm\to \tilde{\chi}_1^0 f f^\prime$ via the $W^\pm$ boson
and the charged Higgs boson $H^\pm$ are studied in the R-parity conserved
Minimal Supersymmetric Standard Model (MSSM).
We treat $\tilde \chi_1^\pm$ decays as production and decay of
$W^\pm$ and $H^\pm$: i.e., $\tilde{\chi}_1^\pm \to  \tilde{\chi}_1^0 W^\pm (H^\pm) \to  \tilde{\chi}_1^0 f f^\prime$.
Both higgsino-like and wino-like $\tilde \chi_1^\pm$ decays are well investigated.
These decays are calculated at 1-loop level and the loop corrections are found to be less than
three percent.
The signal of the charged Higgs $H^\pm$ production from $\tilde \chi_1^\pm$ decays are discussed.
It will offer important information about the chargino and neutralino sector, as well as the charged Higgs sector in the MSSM.

\end{abstract}


\maketitle

\section{Introduction} \label{sec:intro}

The Minimal Supersymmetric Standard Model
(MSSM)\cite{susy, Drees2} is one of the most popular extension of
the Standard Model (SM). Supersymmetry (SUSY) connects fermions with bosons,
which introduces scalar partners to all SM fermions as well as fermionic
partners to all SM bosons. In comparison with SM, two Higgs doublets
are required in the MSSM. After the electroweak symmetry is broken,
it leads to five physical Higgs bosons: three neutral Higgs bosons and
two charged Higgs bosons. Furthermore, superpartners for Higgs bosons and
gauge bosons (the so-called higgsinos and gauginos, respectively)
will mix into charginos and neutralinos, too.
In the R-parity conserved MSSM, the lightest supersymmetric particle (LSP),
which in many scenarios is the lightest neutralino $\tilde{\chi}_1^0$,
appears at the end of the decay chain of each supersymmetric particle. The LSP
escapes the detector, giving the characteristic SUSY signature of
missing energy. Moreover, the stable neutral LSP interacts only
weakly with ordinary matter, it can therefore make a good cold dark
matter candidate.

Heavier supersymmetric particles can be produced at the Large Hadron Collider (LHC)\cite{ATLASTDR},
weakly interaction particles can also be produced at the future $e^+e^-$ collider if
kinematically allowed. Moreover, precision determination of the properties of
supersymmetric particles is expected at future $e^+e^-$ collider\cite{ILC}.
Heavier supersymmetric particles produced at LHC and the future $e^+e^-$ collider
will decay into lighter charginos or neutralinos. Of particular interest are decay chains leading to
the next-to-lightest neutralino $\tilde{\chi}_2^0$ and/or the lighter chargino
$\tilde{\chi}_1^\pm$.
The next-to-lightest neutralino $\tilde \chi_2^0$ in turn can always
decay into the LSP and two SM fermions, which was well studied at 1-loop level \cite{Drees, Neutralino-squark}.
Neutralino decays in the complex MSSM was also studied in ref.\cite{Neutralino-CMSSM}.
Signal of chargino is difficult to be extracted from large $t\bar t$ and $W^+W^-$ backgrounds at the LHC,
while chargino pair production would be easily seen at the future $e^+e^-$ collider due to much more constrained kinematics \cite{Baer1307}.
Depending on the lighter chargino, the lightest neutralino, sfermion as well as charged Higgs boson masses,
the possible decays of the lighter chargino $\tilde \chi_1^\pm$ are three-body
decays $\tilde\chi_1^\pm\rightarrow \tilde \chi_1^0 f f^\prime$,
cascade two-body decays $\tilde\chi_1^\pm\rightarrow \tilde \chi_1^0
W^\pm (H^\pm)\rightarrow \tilde \chi_1^0 f f^\prime$ and
$\tilde\chi_1^\pm\rightarrow \tilde f f^\prime \rightarrow \tilde \chi_1^0 f f^\prime$,
where $f$ and $f^\prime$ are SM fermions. Tree- and one-loop-level chargino decays in
different theory of framework are investigated in refs.\cite{Charginodecayt,Charginodecayl,Charginodecayl2}.
In ref.\cite{Heinemeyer-Complex}, two body decays of chargino to $W$ boson, charged Higgs bosons, as well
sleptons in the complex MSSM are investigated at one-loop level.

In this paper
we concentrate our attention on chargino $\tilde\chi_1^\pm$ decays into $f\bar f^\prime$ via $W$
and charged Higgs boson $H^\pm$ in the real MSSM with heavy sfermions masses.
Suppose the decay mode $\tilde\chi_1^\pm\rightarrow \tilde \chi_1^0 W^\pm$ is open while
others are closed, the branching ratios of decays into $\tilde \chi_1^0 f f^\prime$
are same as $W$ decays to $f f^\prime$ in the SM.  Here we choose SUSY parameters
so that two body decay modes
$\tilde\chi_1^\pm\rightarrow \tilde \chi_1^0 W^\pm\rightarrow \tilde \chi_1^0 f f^\prime$ and
$\tilde\chi_1^\pm\rightarrow \tilde \chi_1^0 H^\pm\rightarrow \tilde \chi_1^0 f f^\prime$ are open, while others are closed kinematically.
The exit of charged Higgs in $\tilde \chi_1^\pm$ decays makes
its branching ratios of decays to $l\nu_l (l=e, \mu, \tau)$ and hadrons final states
are different from that of $W$ decays. This is one of the important signal of the charged Higgs production at the collider, and will offer essential information about the Higgs sector in the MSSM.

This paper is organized as following. In Sec. \ref{sec:2} the MSSM and
the renormalization of those sectors which are relevant
for $\tilde{\chi}_1^\pm$ decays are summarized.  Calculating
techniques are briefly discussed in Sec. \ref{sec:3}. The parameter choices,
numerical results, some  discussions and conclusions are also presented
in this section.

\section{The MSSM and renormalization} \label{sec:2}

In this section we first review the chargino and
neutralino sector, as well as Higgs sector in the R-parity conserved MSSM.
Their renormalization which is required for the precision
calculation is discussed too.

\subsection{Chargino and neutralino sector}

Charginos and neutralinos are mixture of charged and neutral
gauginos and higgsinos, respectively. In the gaugino and higgsino
eigenbasis, the mass terms of the charginos and neutralinos can be
written as
\bea -\mathcal{L}_{\chi^c-{\rm mass}}  &=&  \psi_R^{T} X \psi_{L} +
h.c. \, ,\nonumber \\
-\mathcal{L}_{\chi^0-\rm mass}  &=&  \frac{1}{2} {\psi^{0}}^T Y
\psi^{0} + h.c. \eea Here $\psi_{L}$,$\psi_{R}$ and $\psi^0$ are
two-components Weyl spinors, their expressions are shown as
following,
\beq \psi_{L} =
(\widetilde{W}^{+},~~\tilde{h}^{+}_{2})^{T},~~ \psi_{R} =
(\widetilde{W}^{-},~~\tilde{h}^{-}_{1})^{T},~~ \psi^{0} =\left(
\begin{array}{cccc} \tilde{B^{0}}, & \widetilde W^3, &
    \tilde h_1^0, & \tilde h_2^0 \end{array}\right)^{T}.
 \eeq
 The mass matrices for charginos and neutralinos are
\beq
\label{mc}
X= \left( \begin{array}{cc}
M_{2} & \sqrt{2} M_W s_\beta \\
\sqrt{2} M_W c_\beta & \mu \\
\end{array} \right),
\eeq
 \beq \label{mn} Y=\left( \begin{array}{llll}
M_{1} & 0 & -M_{Z}s_W c_\beta & M_{Z}s_W s_\beta\\
0 &  M_{2} & M_{Z}c_W c_\beta & -M_{Z}c_W s_\beta \\
-M_{Z}s_W c_\beta & M_{Z}c_W c_\beta & 0 & -\mu \\
M_{Z}s_W s_\beta & -M_{Z}c_W s_\beta & -\mu & 0 \end{array}
\right)\,.
\eeq
Here $M_1$ is the SUSY breaking $U(1)_Y$ gaugino (bino) mass, $M_2$
is the SUSY breaking $SU(2)$ gaugino (wino) mass, $\mu$ is the
supersymmetric higgsino mass, and $\tan\beta$ is the ratio of
vacuum expectation values of
the two neutral Higgs fields of the MSSM. Abbreviations $s_W,
s_\beta, c_W$ and $c_\beta$ stand for $\sin \theta_W, \sin \beta,
\cos \theta_W$ and $\cos \beta$, respectively, where $\theta_W$ is
the weak mixing angle.

Mass matrices $X$ and $Y$ can be diagonalized by transforming the
original wino and higgsino fields with the help of unitary matrices,
\beq \chi_L = V {\psi}_L,~~ \chi_R = U \psi_R\,,~~ M_{\tilde \chi^+}
= UXV^T = {\rm diag}\left (m_{\tilde \chi^+_1}, m_{\tilde
\chi^+_2}\right );\label{char-rotat}
 \eeq
\beq \chi^0 = N\psi^0,~~M_{\tilde \chi^0} = N^*YN^\dag = {\rm
diag}\left (m_{\tilde \chi^0_1}, m_{\tilde \chi^0_2}, m_{\tilde
\chi^0_3}, m_{\tilde \chi^0_4}\right ). \label{neu-rotat}\eeq Here
$U,V$ are unitary $2\times 2$ matrices which determined by the third
part of eq.(\ref{char-rotat}), $N$ is a unitary $4\times 4$ matrix
which determined by the second part of eq.(\ref{neu-rotat}), $
\chi_{L/R}, \chi^0$ are chargino and neutralino mass eigenstates,
respectively. The four-component spinors for chargino and neutralino
are defined by \beq \tilde \chi_i^+ = \left( \begin{array}{c}
{\chi_L}_i\\
\overline {\chi_R}_i
\end{array} \right),~~~~~
\tilde \chi_i^0= \left(
\begin{array}{c}
\chi_i^0\\
\overline {\chi_i^0}
\end{array} \right),
\eeq
where neutralinos are Majorana fermions. There are two
charginos and four neutralinos in the MSSM. They are labeled in
ascending order,
\beq
0 < m_{\tilde \chi_{1}^{+}} \leq m_{\tilde
\chi_{2}^{+}},~~ 0 \leq m_{\tilde{\chi}^{0}_{1}} \leq
m_{\tilde{\chi}^{0}_{2}} \leq m_{\tilde{\chi}^{0}_{3}} \leq
m_{\tilde{\chi}^{0}_{4}}.
\eeq
In the R-parity conserved MSSM, the
lightest neutralino $\tilde{\chi}^{0}_{1}$ can be a good cold dark
matter candidate.

Concerning the renormalization of chargino and neutralino sector at
one-loop level, different approaches were developed
\cite{Drees,Fritzsche,Oller, Eberl,Guasch, Chatterjee, Fritzsche:2011nr, ReNeutalinoChargino}.
Here we assume all the parameters are real and employ
the on-shell renormalization following refs.\cite{Drees, Fritzsche,
Chatterjee}. Mass matrices and fields of charginos and neutralinos
are renormalized as following
\begin{eqnarray}
X &\longrightarrow & X + \delta X \, ,~~Y \longrightarrow  Y +
\delta Y , \label{eqn:ReCha1a}  \\
\omega_{L}\tilde{\chi}_i & \longrightarrow & \left(\delta_{ij} +
  \frac{1}{2}\left (\delta Z^L\right )_{ij}\right)
\omega_{L}\tilde{\chi}_j\, ,
\nonumber \\
\omega_{R}\tilde{\chi}_i& \longrightarrow &\left(\delta_{ij} +
  \frac{1}{2}\left (\delta Z^R\right )^{\ast}_{ij}\right)
\omega_{R}\tilde{\chi}_j\, , \label{eqn:ReCha1b}
\end{eqnarray}
where each element of $\delta X$ and $\delta Y$ is the counterterm
for the corresponding entry in $X$ and $Y$ mass matrices,
respectively. In eq.(\ref{eqn:ReCha1b}), $\omega_{L,R}=(1 \mp
\gamma_5)/2$ are chiral operators and this equation holds for both
charginos, with $\tilde \chi_i \equiv \tilde \chi_i^+, \, i \in
\{1,2\}$, and neutralinos, with $\tilde \chi_i \equiv \tilde
\chi_i^0, \, i \in \{1,2,3,4\}$. Note that the right- and
left-handed field renormalization constant for neutralinos are same,
i.e. $\delta Z^L = \delta Z^R = \delta Z^0$, since they are Majorana
fermions.

Altogether mass counterterms $\delta X$ and $\delta Y$ contain
seven different counterterms: $\delta M_W$, $\delta M_Z$, $\delta
\theta_W$, $\delta \tan\beta$, $\delta M_1$, $\delta M_2$ and
$\delta \mu$. The first three of these already appear in the SM and
their renormalization have been discussed in ref.\cite{SMDenner},
we will not repeat it here. Parameter $\tan\beta$ will be renormalizated
in Higgs sector in the next subsection.
In the on-shell renormalization scheme
for the charginos/neutralinos \cite{Fritzsche,Chatterjee}, the
counterterms $\delta M_1$, $\delta M_2$ and $\delta \mu$ are
determined by requiring that three pole-masses of six chargnios and
neutralinos are the same as at tree-level. Ref.\cite{Chatterjee} has
studied all instabilities and singularities of different type of
choices for inputs.
It concludes that one should choose the masses of a bino-like, a
wino-like and a higgsino-like state as inputs in order to avoid
large corrections to the masses of the other eigenstates. In this
paper, We keep masses of $\tilde \chi_1^0, \, \tilde \chi_1^+$ and
$\tilde \chi_2^+$ unchanged at tree- and one-loop-level, as in
ref.\cite{Fritzsche}. In our numerical set-up, see Sec.\ref{sec:3},
the lightest neutralino is always bino-like. This makes our choices
reasonable. Considering the on-shell field renormalization of
charginos and neutralinos, the diagonal entries of the field
renormalization constants are fixed by the condition that the
corresponding renormalized propagator has unit residue. Furthermore,
the renormalized one-particle irreducible two-point functions should
be diagonal for on-shell external particles, which fixes the
off-diagonal entries of the field renormalization constants.

\subsection{Charged Higgs sector}\label{subsec:Higgs}
The mass term for the charged Higgs at tree level can be expressed
as
\beq
\mathcal{L}_{H^\pm_{\rm mass}} =\left (\begin{array}{c}
 H^+, G^+\end{array}\right)\left (\begin{array}{c}
 m_{H^\pm}^2, m_{H^+G^-}^2\\
 m_{H^-G^+}^2,m_{G^\pm}^2\end{array}\right)\left (\begin{array}{c}  H^-\\
G^- \end{array}\right)\, .\label{eqn:ChargeHiggs}
\eeq
The mass matrix elements are as following,
\begin{eqnarray}
 m_{H^\pm}^2 & = & m_{A^0}^2 + M_W^2\, ,\nonumber\\
m_{H^+G^-}^2&=&m_{H^-G^+}^2=-\left ( m_{A^0}^2 + m_W^2\right )\tan(\beta - \beta_c) \,  \nonumber \\
&&{}- \frac{e}{2m_Z s_W c_W}\left (T_{H^0}\sin(\alpha - \beta_c) +T_{h^0}\cos(\alpha - \beta_c)\right ) /\cos(\beta - \beta_c)\, , \nonumber\\
 m_{G^\pm}^2& = & \left ( m_{A^0}^2 + m_W^2\right )\tan^2(\beta - \beta_c)\,  \nonumber \\
&&{}- \frac{e}{2m_Z s_W c_W} T_{h^0}\cos(\alpha + \beta - 2\beta_c) /\cos^2(\beta - \beta_c)\,  \nonumber \\
&& {} + \frac{e}{2m_Z s_W c_W} T_{H^0}\sin(\alpha + \beta -
2\beta_c) /\cos^2(\beta - \beta_c)\, . \label{eqn:ChargeHiggsmass}
\end{eqnarray}
Here $m_{A^0}$ is the mass for the neutral CP-odd Higgs boson
$A^0$,$\alpha$ is the mixing angle of two neutral CP-even Higgs
bosons, $\beta_c$ is the mixing angle of two charged Higgs bosons.
$T_{h^0}$ and $T_{H^0}$, denote tadpoles of the physical neutral Higgs
fields $h^0$ and $H^0$, are zero at tree-level. The mass matrix in
eq.(\ref{eqn:ChargeHiggs}) should be diagonal at tree level.
This leads to the following conclusions,
\begin{eqnarray}
\beta_c  = \beta, \hspace{5mm} \tan2\alpha = \tan2\beta \frac{m_{A^0}^2
+ m_Z^2}{m_{A^0}^2 - m_Z^2}, \hspace{5mm}-\frac{\pi}{2}<\alpha
<\frac{\pi}{2}\, .
\end{eqnarray}

Concerning renormalization of the Higgs sector, we follow the approach
in Ref.\cite{ReHiggs}.
Introduce renomalization constants for the mass matrix and fields of
the charged Higgs sector as following, \begin{eqnarray} m_{H^\pm}^2
&\rightarrow& m_{H^\pm}^2 +\delta m_{H^\pm}^2\, ,\\
m_{H^-G^+}^2 &\rightarrow& m_{H^-G^+}^2 + \delta m_{H^-G^+}^2\, ,\\
m_{G^\pm}^2 &\rightarrow& m_{G^\pm}^2 + \delta m_{G^\pm}^2\, ,\\
\left (\begin{array}{c}  H^-\\
G^- \end{array}\right) &\rightarrow& \left (1+ \frac{1}{2}\delta Z \right )\left (\begin{array}{c}  H^-\\
G^- \end{array}\right)\, .
\end{eqnarray}
Here the filed renormalization constant $\delta Z$ is a $2\times 2$
matrix, which is fixed by using $\overline{DR}$ scheme, which means
that the counterterms only contain UV-divergent parts. The mass
counterterms $\delta m_{H^\pm}^2$, $\delta m_{H^-G^+}^2$ and $\delta
m_{G^\pm}^2$ contain counterterms $\delta T_{h^0}, \delta T_{H^0}$,
$\delta m_{A^0}^2$, and $\delta \tan\beta$. The counterterm $\delta
m_{A^0}^2$ is determined by renormalizing the neutral CP-odd Higgs
boson $A^0$ via the on-shell renormalization scheme. Counterterms
for the tadpoles $T_{h^0}$ and $T_{H^0}$ are fixed by requiring the
renormalized tadpoles are equal to zero at one-loop level. Same as
the field renormalization constants of the charged Higgs, $\delta
\tan\beta$ is determined in the $\overline{DR}$
scheme\cite{RetanBMS},
\begin{eqnarray}
\frac{\delta\tan\beta^{\overline{\rm DR}}}{\tan\beta}& = &
\frac{1}{2m_Zs_\beta c_\beta} \bigl[ \mathrm{Im}
\Sigma_{A^0Z}(m_A^2) \bigr]_{\rm div}\, .
 \end{eqnarray}
Here the subscript "div' means that only the UV-divergent parts are
considered. This scheme has the advantage of providing the gauge
invariant and process independent counterterms.

\section{Calculations and numerical results} \label{sec:3}

In this work SUSY parameters are chosen to make the
 cascade two-body decays of lighter chargino $\tilde \chi_1^\pm$ via
 $W^\pm$ and charged Higgs boson $H^\pm$ possible. No other two-body  decay mode is open. The decays $\tilde{\chi}_1^\pm \rightarrow \tilde{\chi}_1^0
f \bar f^\prime$ can be approximately treated as production and
decays of the $W^\pm$ and charged Higgs boson $H^\pm$,
\bea
\Gamma(\tilde{\chi}_1^\pm \rightarrow  \tilde{\chi}_1^0 f \bar f^\prime)
&\simeq &
\Gamma(\tilde{\chi}_1^\pm \rightarrow \tilde{\chi}_1^0W^\pm)
 Br(W^\pm\rightarrow f \bar f^\prime) \non
 && + \Gamma(\tilde{\chi}_1^\pm \rightarrow \tilde{\chi}_1^0H^\pm)
 Br(H^\pm\rightarrow f \bar f^\prime), \\
Br(\tilde{\chi}_1^\pm \rightarrow  \tilde{\chi}_1^0 f \bar f^\prime)
 &\simeq &
Br(\tilde{\chi}_1^\pm \rightarrow \tilde{\chi}_1^0W^\pm)
 Br(W^\pm\rightarrow f \bar f^\prime) \non
 && + Br(\tilde{\chi}_1^\pm \rightarrow \tilde{\chi}_1^0H^\pm)
 Br(H^\pm\rightarrow f \bar f^\prime),
\eea
where $Br(W^\pm\rightarrow f \bar f^\prime)$  and $Br(H^\pm\rightarrow
 f \bar f^\prime)$ are branching ratios of $W^\pm$ and $H^\pm$ boson decay to
two SM fermions, respectively. Since the branching ratios of
$W$ boson decays have been
measured precisely, we here will not repeat the theoretical calculation, but
take the measured values from Particle Data Group
\cite{pdg2012}. For the relevant charged Higgs decays in the MSSM,
they are calculated at one-loop level by using the program {\sc FeynHiggs}\cite{FeynHiggs}.
Suppose the couplings of charged Higgs with fermions are well known, one can determine the branching ratios
of $\tilde \chi_1^- \to \tilde \chi_1^0 W^- (H^-)$ from the measured branching ratios
$Br\left(\tilde \chi_1^- \to \tilde \chi_1^0 \tau^- \bar \nu_\tau \right )$ by
\beq
Br\left(\tilde \chi_1^- \to \tilde \chi_1^0 \tau^- \bar \nu_\tau \right) = x Br\left(W^- \to \tau^- \bar\nu_\tau \right)+
 (1-x) Br \left(H^- \to \tau^- \bar \nu_\tau\right),
\eeq
where $x$ and $1-x$ are the branching ratios of $\tilde \chi_1^-$ decays to $W^-$ and $H^-$, respectively.

Decays $\tilde{\chi}_1^\pm \rightarrow \tilde{\chi}_1^0W^\pm$ and
$\tilde{\chi}_1^\pm \rightarrow \tilde{\chi}_1^0H^\pm$ are
calculated at one-loop level with the help of the packages {\sc
FeynArts}\cite{FeynArts}, {\sc FormCalc} and {\sc LoopTools} \cite{FormCLT},
respectively. The virtual contributions of these processes only
contain vertex type corrections, which are ultraviolet(UV)
divergent. These corrections become UV-finite after adding the
contributions of the counterterms that originate from the
renormalization of the MSSM, as discussed in Sec.~\ref{sec:2}.
Virtual diagrams with photon attached to two external particles will
give infrared (IR) divergences, which are regularized by a photon
mass. When the photon energy $E_\gamma$ is very small, the real
photon bremsstrahlung will also give IR-divergent contribution which
is sufficient to cancel the IR divergences from the virtual
corrections. Contribution of the real photon bremsstrahlung is
split into two parts: the ``soft photon
bremsstrahlung''($E_{\gamma}\le \Delta E$) and the ``hard photon
bremsstrahlung''($E_{\gamma}> \Delta E$) contribution, here the
cutoff parameter $\Delta E$ should be small compared to the relevant
physical energy scale. The contribution of the soft photon
bremsstrahlung can be described as a convolution of the differential
tree-level decay width with a universal factor. Explicit expressions
can be found in Refs. \cite{SMDenner, soft}. Since external charged
particles in processes $\tilde{\chi}_1^\pm \rightarrow
\tilde{\chi}_1^0W^\pm/H^\pm$ are quite heavy, contribution of the
hard photon bremsstrahlung contains no collinear divergences and can be
calculated numerically using Monte Carlo integration. The dependence
on the largely arbitrary parameter $\Delta E$ cancels after summing
soft and hard contributions, provided it is sufficiently small.

Considering the constraint on SUSY parameters from
recent experiments \cite{LHC-SUSY}, the soft SUSY-breaking parameters
in the diagonal entries of the sfermion mass matrices are chosen to
be the same
\bea
M_{SUSY} = 1.5TeV\, , \label{eq:msusy}
\eea
while the trilinear couplings of the third generation and other relevant
input parameters are chosen as
\bea
A_t = A_b = A_{\tau} = 2.5 {\rm TeV}, \quad
M_1 = 150 GeV, \quad M_{H^\pm} = 160 GeV,  \quad M_{\tilde g} = 1 {\rm TeV}.
\label{eq:inputs}
\eea
As discussed in Sec.~\ref{sec:2}, pole masses
of the lightest neutralino and two charginos are chosen to be input
parameters in our on-shell renormalization approach. Parameters $M_2$ and
$\mu$ therefore can be expressed as a function of pole masses of two charginos,
see Ref.\cite{Fritzsche}. For given pole masses of two charginos, there
are two type of choices for parameters $M_2$ and $\mu$ \cite{Heinemeyer-Complex}: $M_2>\mu$
and $M_2 < \mu$, which make the lighter chargino $\tilde \chi_1^\pm$ is more
higgsino-like and gaugino-like, respectively.
Though small $\mu$ is preferred in Natural SUSY\cite{NaturalSUSY},
here we focus on more general cases. In our calculation, parameters $M_2$ and $\mu$ are chosen as
in Table \ref{choice1}, where $\tan\beta = 20$.
\begin{table}[htb] \label{spstab}
\begin{center}
\caption{Different choices for parameters $M_2$ and $\mu$ for $\tan\beta = 20$:
Set-I to Set-IV. \label{choice1} }
\vspace{0.5cm}
\begin{tabular}{|c|c|c|c|c|}\hline
& \multicolumn{2}{|c|}{$M_2>\mu$} & \multicolumn{2}{|c|}{$M_2<\mu$} \\
\cline{2-5}
& Set-I & Set-II & Set-III &Set-IV \\ \hline
$M_2$ (GeV)& $600$ &$550-800$ &  $320$ & $320-550$ \\ \hline
$\mu$ (GeV)& $320-550$ & $320$ & $550-800$ & $600$ \\ \hline
\end{tabular}
\end{center}
\end{table}

\begin{figure}[b!]
\begin{tabular}{cc}
\psfrag{Br}{$B_r$}
\psfrag{M2}{$M_2$}
\includegraphics[width=0.5\linewidth]{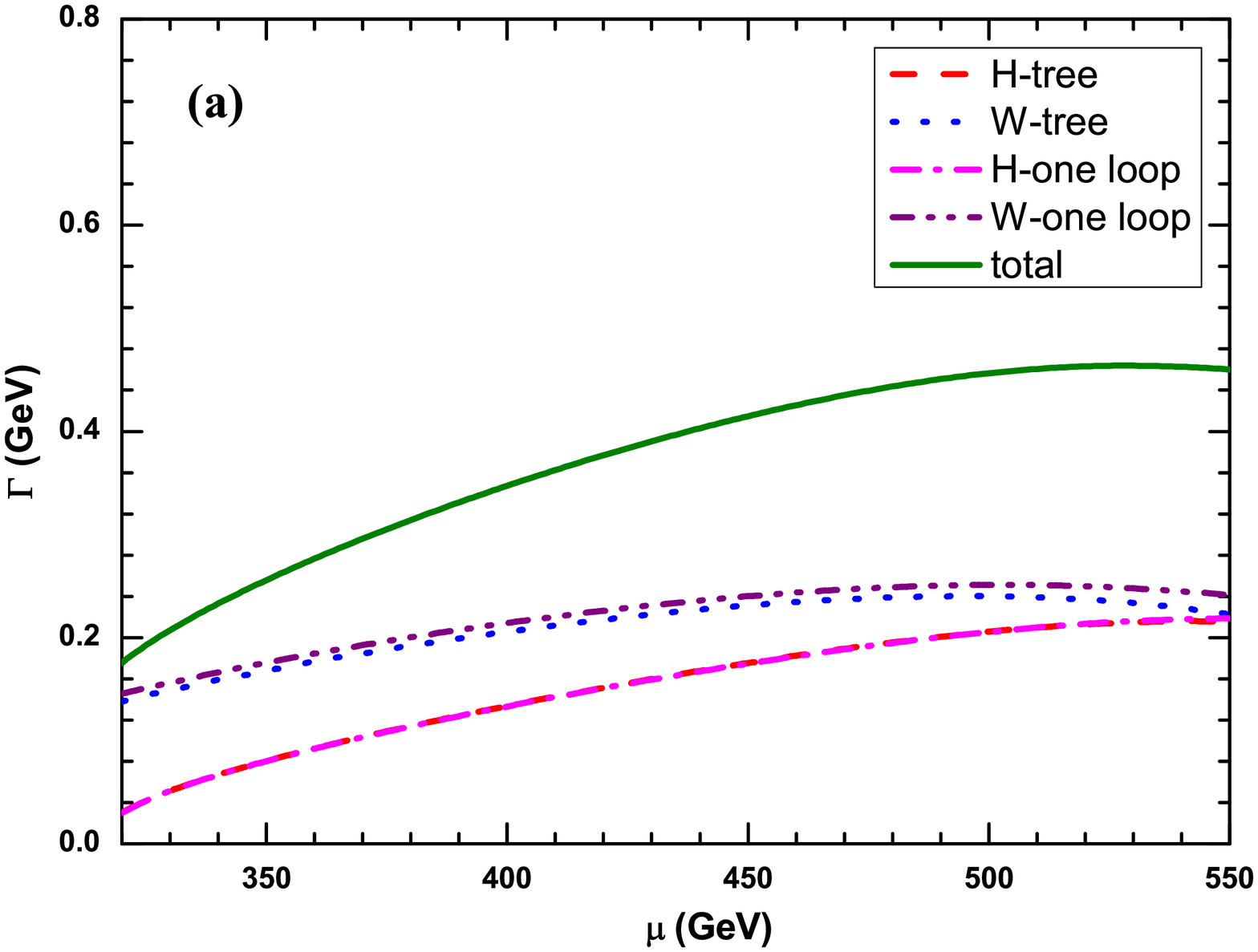}&
\includegraphics[width=0.5\linewidth]{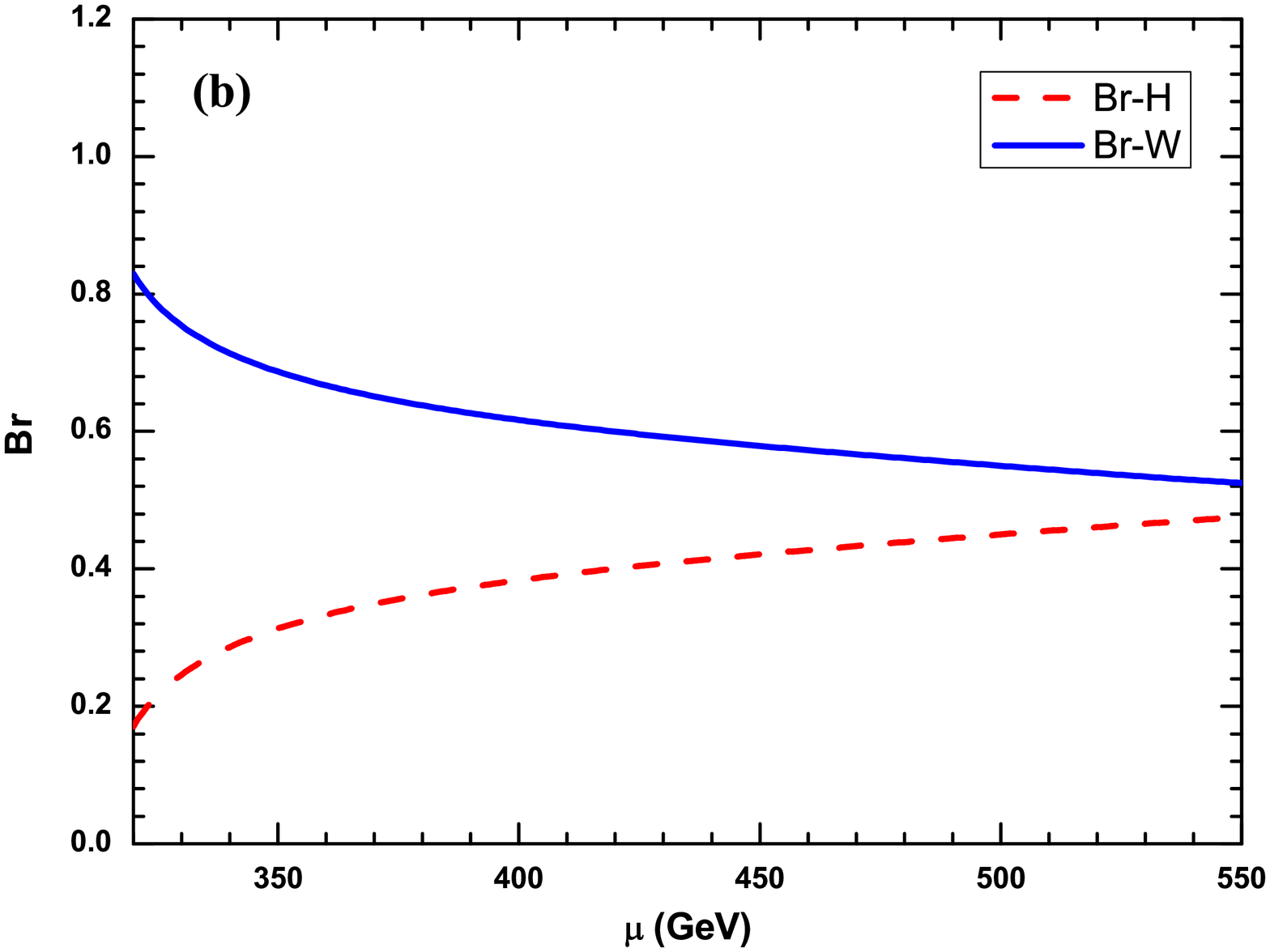} \\
\includegraphics[width=0.5\linewidth]{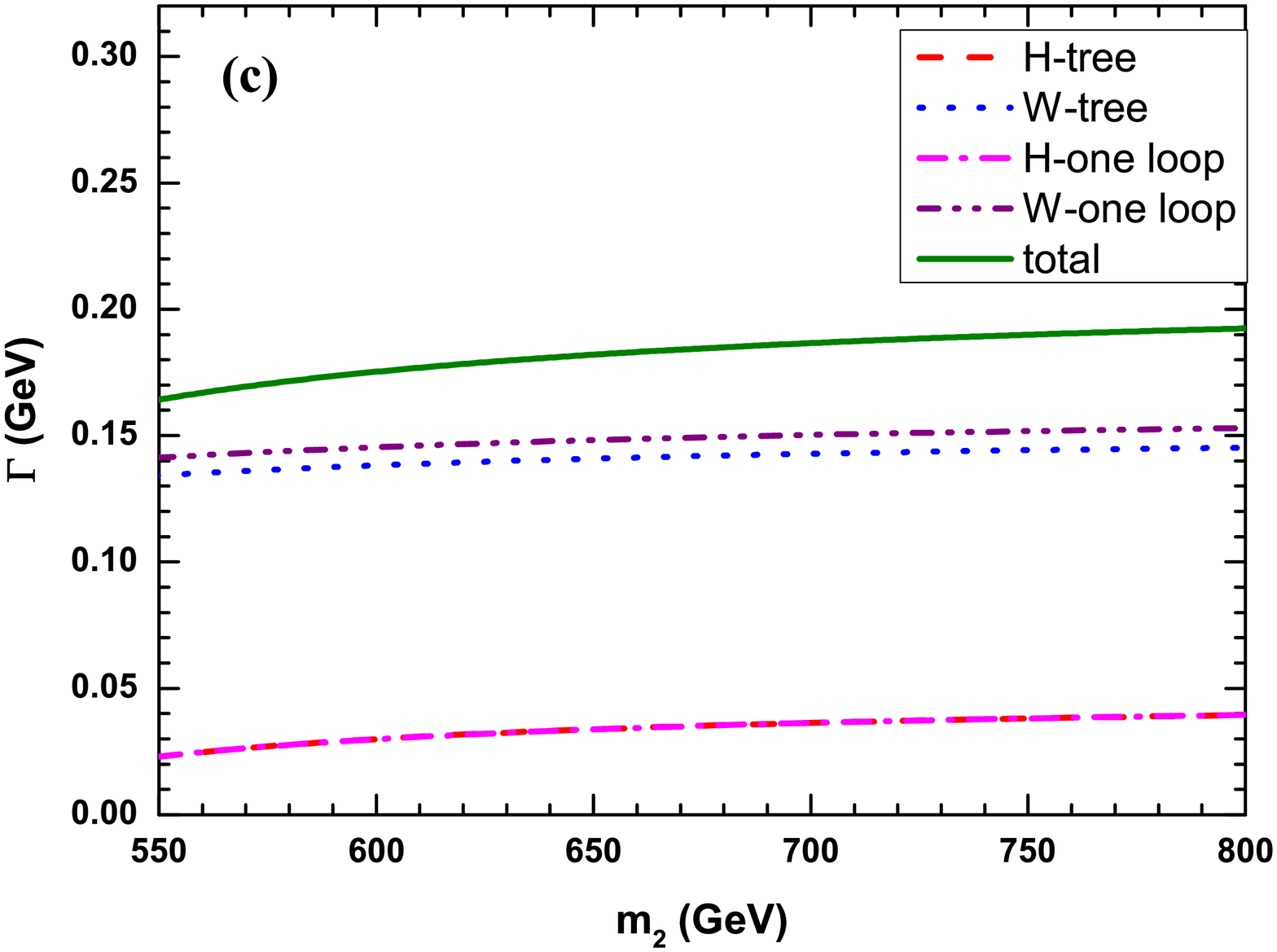} &
\includegraphics[width=0.5\linewidth]{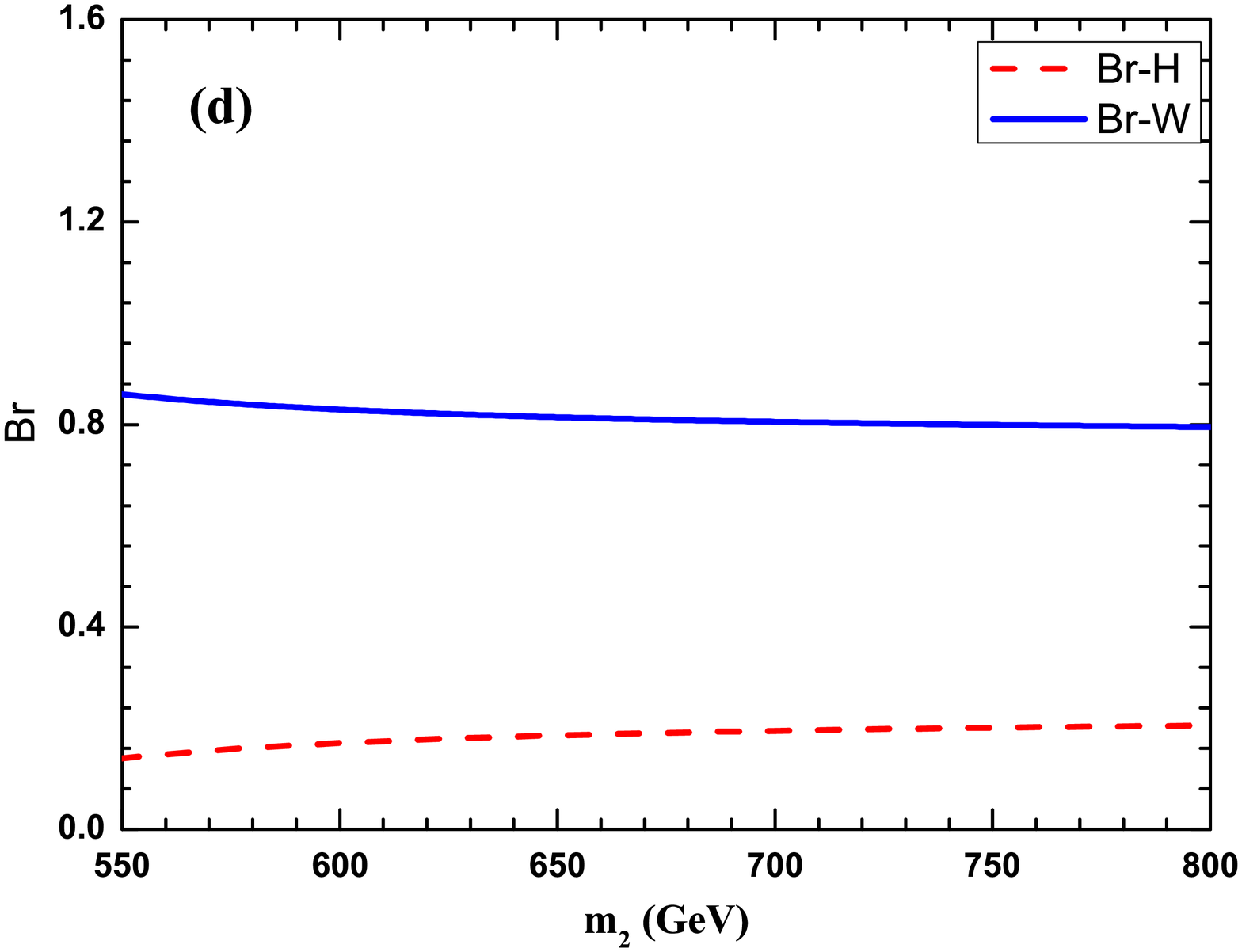}
\end{tabular}
\caption{ The decay widths and branching ratios of
$\tilde \chi_1^- \to \tilde \chi_1^0 W^- (H^-)$ decays
for the case of $M_2 > \mu$.
The curves in (a)-(b)( (c)-(d)) are obtained by using the Set-I (Set-II) input parameters
as defined in Table I.\label{fig:1} }
\end{figure}

Note that all SUSY parameters as given in Eqs.(\ref{eq:msusy}-\ref{eq:inputs}) and in Table I are real numbers.
It ensures CP is conserved in our consideration, ie. the decay rate of $\tilde \chi_1^-$ is
exactly same as its conjugate state $\tilde \chi_1^+$.
In this paper only $\tilde \chi_1^-$ decay is investigated, the conclusion can be applied to
$\tilde \chi_1^+$ decay.
By using the fixed input parameters as given in Eqs.(\ref{eq:msusy}-\ref{eq:inputs}) and
different choices for $M_2$ and $\mu$ as listed in Table I, we calculate the decay widths and
branching ratios for all considered decays, and show the theoretical predictions in Figs.1-3.

In Fig.1 the decay widths and branching ratios of the lighter
chargino decays to the lightest neutralino and $W^-/H^-$ boson at the tree and one-loop level
are presented for the case of $M_2>\mu$, i.e. $\tilde \chi_1^-$ is higgsino-like.
In Fig.1(a)-1(b), we show the $\mu$-dependence of the
decay widths and branching ratios of $\tilde \chi_1^- \to \tilde \chi_1^0 W^- (H^-)$
decays. In Fig.1(c)-1(d), we show the $M_2$-dependence of the
decay widths and branching ratios of the same decay modes.
From the curves in these figures one can see the following points:
\begin{itemize}
\item
For the parameter choice Set-I, the mass of $\tilde \chi_1^-$ increases with $\mu$ which
makes decay width of $\tilde \chi_1^- \to \tilde \chi_1^0 W^-/H^-$ and total decay width of $\tilde \chi_1^-$ increase
kinematically, see Fig.1(a)-1(b).
Couplings of $\tilde \chi_1^- \tilde \chi_1^0 W^+$ (higgsino-higgsino, wino-wino interaction) and
$\tilde \chi_1^- \tilde \chi_1^0 H^+$ (wino-higgsino, higgsino-wino, higgsino-bino interaction)
vary with $\mu$, too. The competition of kinematics and couplings makes the branching ratio
$Br(\tilde \chi_1^- \to \tilde \chi_1^0 H^-)$ increases while
$Br(\tilde \chi_1^- \to \tilde \chi_1^0 W^-)$
decreases when $\mu$ becomes large, as illustrated by the curves in Fig.1(a) and 1(b).
\item
For the parameter choice Set-II, $M_2$ is much bigger than $\mu$.
Increasing $M_2$ will not change the higgsino and gaugino part of the
lighter chargino $\tilde \chi_1^-$ too much and
hence the couplings of $\tilde \chi_1^- \tilde \chi_1^0 W^+(H^+)$
are nearly invariant. On the other hand, the mass of
$\tilde \chi_1^-$ depends mainly on the parameter $\mu$ in case of
$M_2>\mu$, varying parameter $M_2$ make little difference on
the kinematics. This makes the decay width and branching ratios
of $\tilde \chi_1^-$ change
a little bit with varying parameter $M_2$ once the
decay channel $\tilde \chi_1^- \to \tilde \chi_1^0 H^-$ is open, see Fig.1(c) and 1(d).
\end{itemize}

\begin{figure}[htb]
\begin{tabular}{cc}
\includegraphics[width=0.5\linewidth]{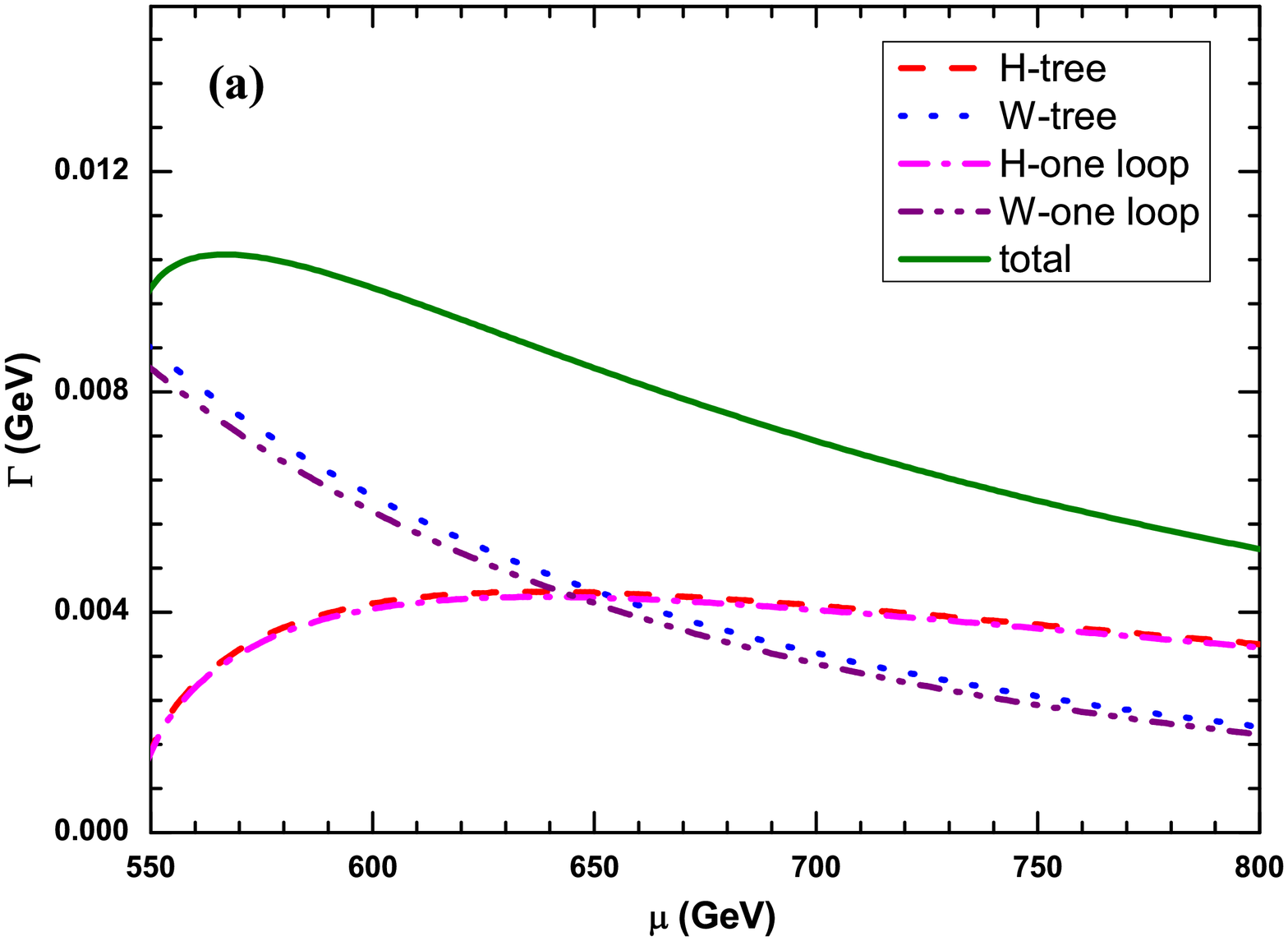}&
\includegraphics[width=0.5\linewidth]{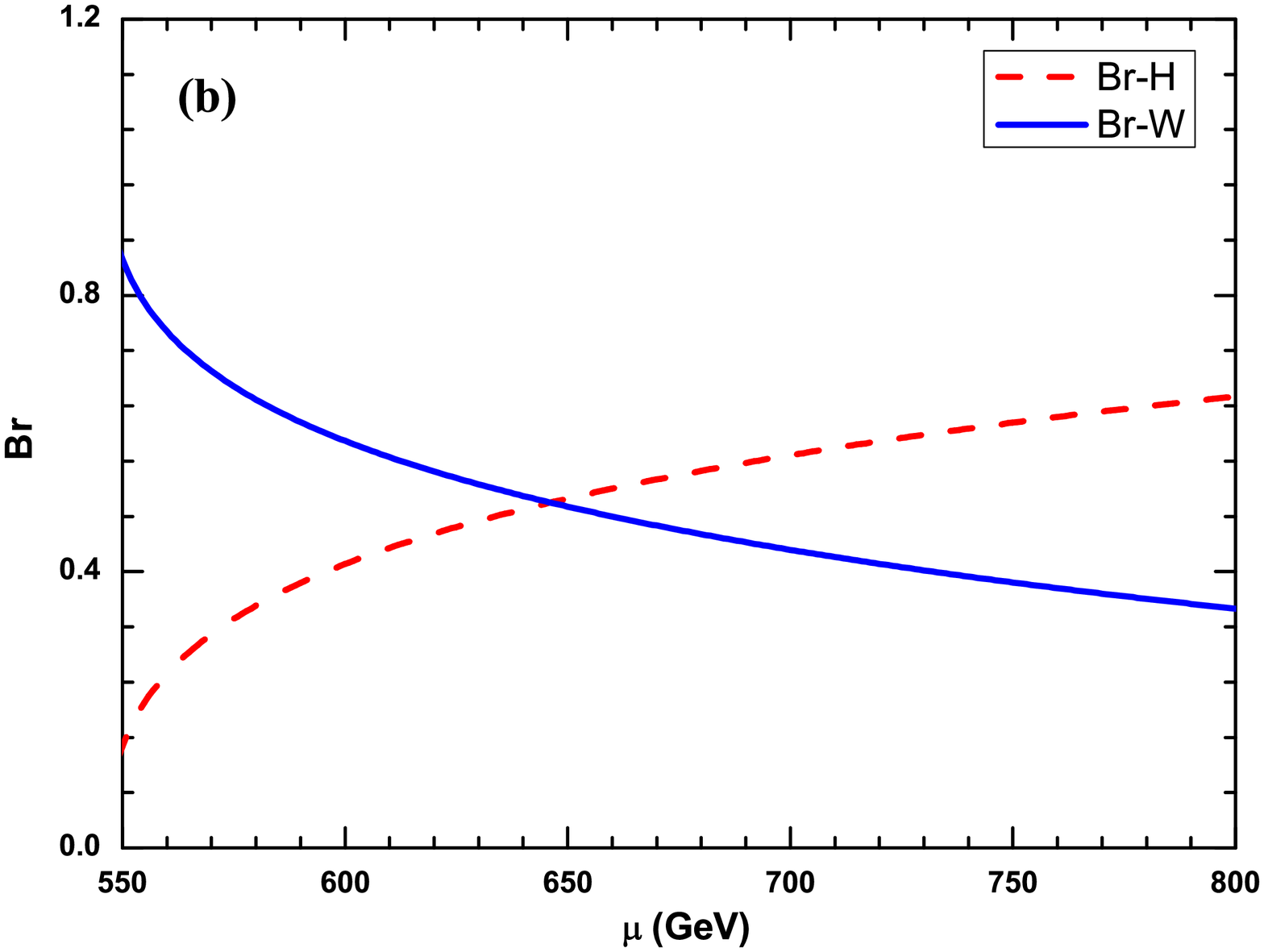}\\
\includegraphics[width=0.5\linewidth]{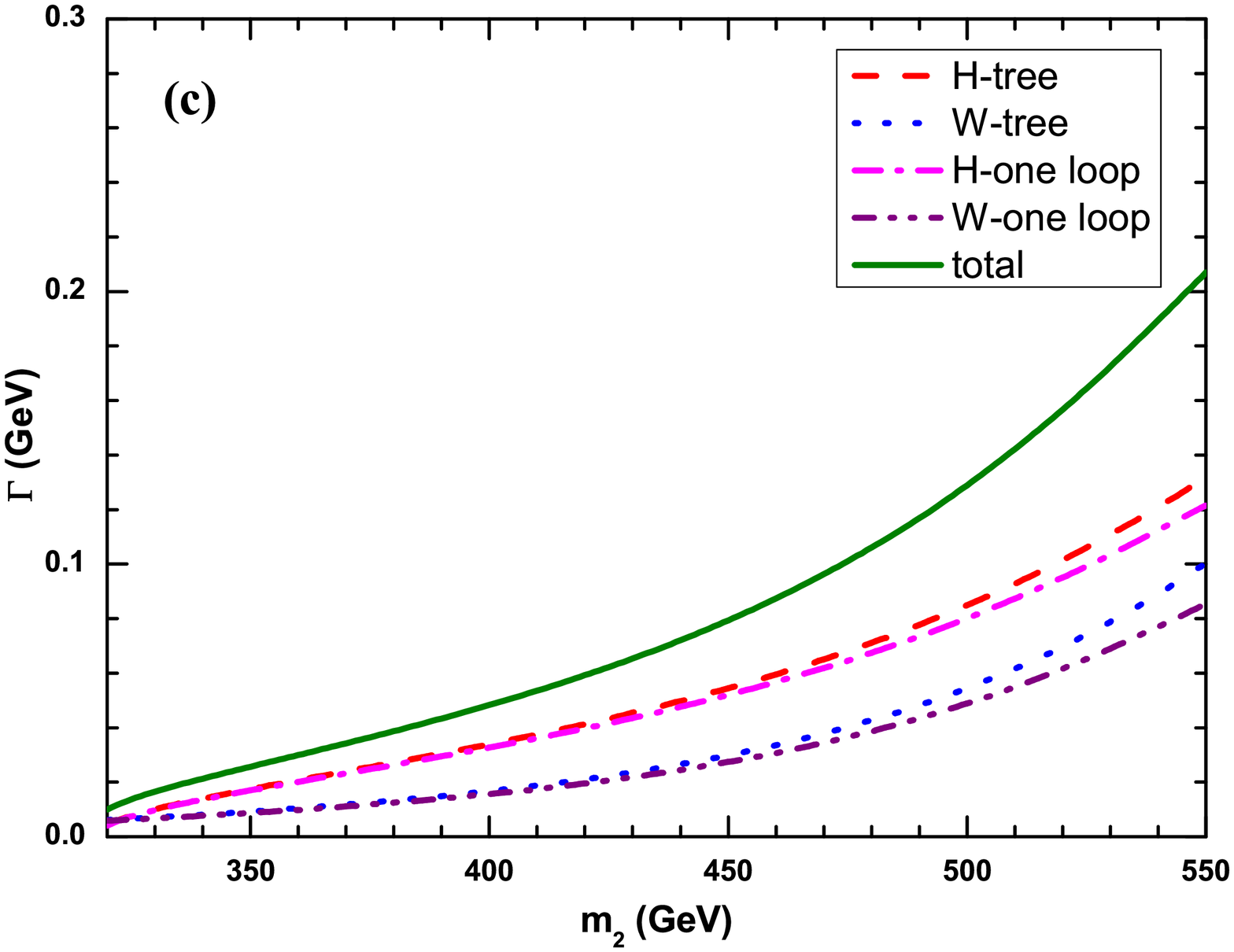}&
\includegraphics[width=0.5\linewidth]{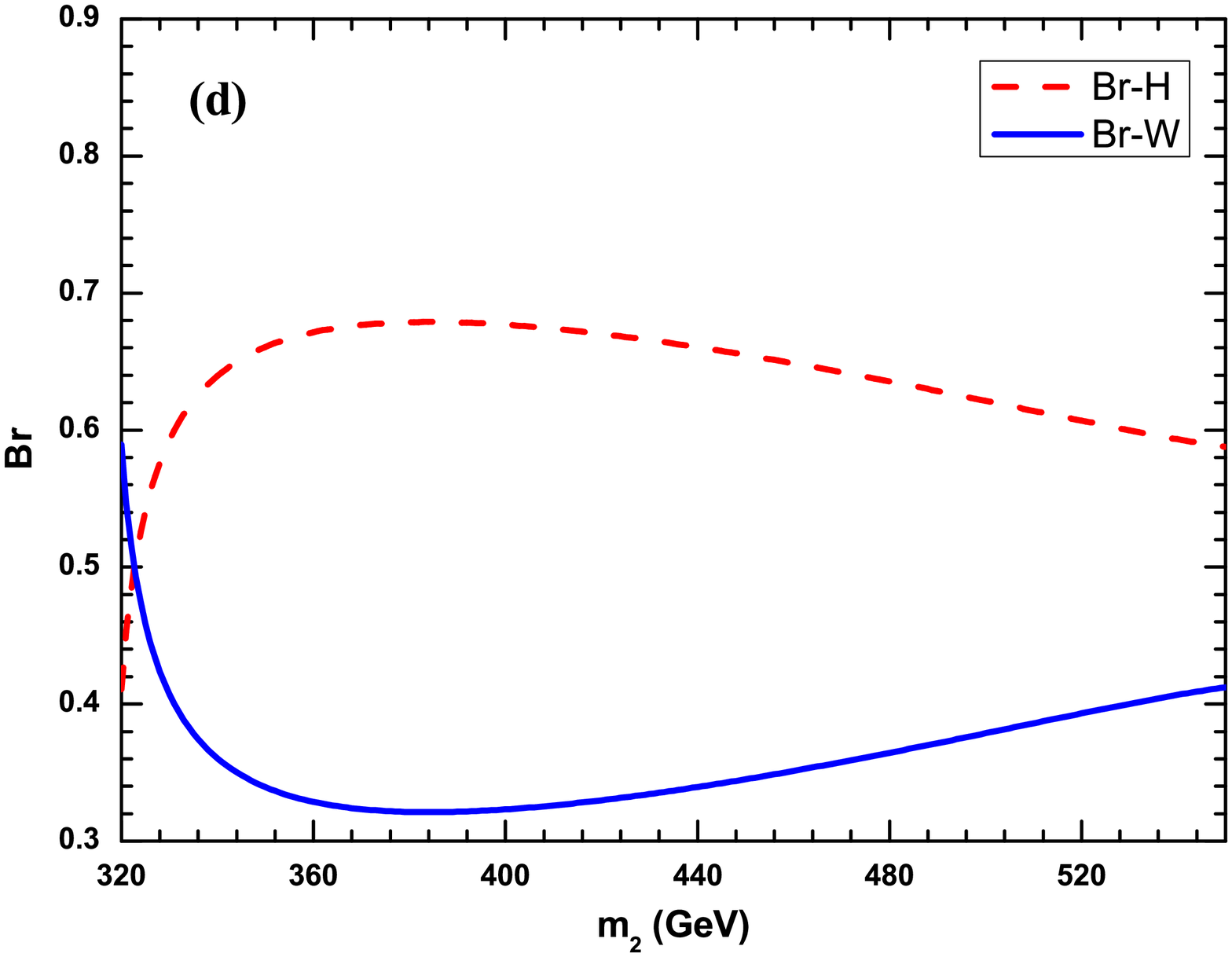}
\end{tabular}
\caption{The decay widths and branching ratios
of $\tilde \chi_1^- \to \tilde \chi_1^0 W^- (H^-)$
for $M_2 < \mu$. The curves in (a) and (b) ( (c) and (d) ) are calculated
by using  the Set-III ( Set-IV) input parameters as defined in Table I.\label{fig:2}}
\end{figure}

In Fig.\ref{fig:2} the decay widths and branching ratios of
the $\tilde \chi_1^- \to \tilde \chi_1^0 W^- (H^-)$ decays
at the tree and
one-loop level are illustrated for the case of $M_2 < \mu$, i.e.
$\tilde \chi_1^-$ is wino-like.
\begin{itemize}
\item
In Fig.2(a)-2(b), the theoretical predictions are obtained by using the parameter choice
Set-III: $M_2=320GeV, \mu=550GeV\sim 800GeV$.
The effects of 2-body kinematics on the decay width is quite small.
Couplings of $\tilde \chi_1^- \tilde \chi_1^0 W^+ (H^+)$ decrease
with increasing parameter $\mu$, but the descent speed for coupling
$\tilde \chi_1^- \tilde \chi_1^0 W^+$ is much faster than that of
couplings $\tilde \chi_1^- \tilde \chi_1^0 H^+$.
The branching ratio of $\tilde \chi_1^- \to \tilde \chi_1^0 W^-$
($\tilde \chi_1^- \to \tilde \chi_1^0 H^-$ ) become consequently smaller (larger)
when the scale $\mu$ increases, as shown by the curves in Fig.2(a) and 2(b).

\item
In Fig.2(c)-2(d), we show the theoretical predictions obtained by using the
parameter choice Set-IV: $M_2=320\sim 550GeV, \mu=600GeV$
The competition of the kinematics and couplings makes
the decay widths of two decay modes increase with increasing $M_2$, and
the branching ratio $Br(\tilde \chi_1^- \to \tilde \chi_1^0 H^-)$
is larger than $Br(\tilde \chi_1^- \to \tilde \chi_1^0 W^-)$ in
almost all the parameter space.
\end{itemize}
From Figs.1 and 2 one can see that the loop effects on the decay widths
are very small: less than $3\%$ in magnitude.
Furthermore, comparing Fig.1(a)-1(b) with Fig.2(c)-2(d) (Fig.1(c)-1(d) with Fig.2(a)-2(b)),
one concludes that exchanging values for parameter $M_2$ and $\mu$ will change the decay width
of charginos and its branching ratios, though their masses are fixed.
The branching ratio $Br(\tilde \chi_1^- \to \tilde \chi_1^0 H^\pm)$ is smaller than $Br(\tilde \chi_1^- \to \tilde \chi_1^0 W^-)$
for $M_2>\mu$, while $Br(\tilde \chi_1^- \to \tilde \chi_1^0 H^\pm)$ can be comparable with or even larger than $Br(\tilde \chi_1^- \to \tilde \chi_1^0 W^-)$
for $M_2<\mu$, for ours choice of parameters in Table I.

\begin{figure}[tb]
\begin{tabular}{cc}
\psfrag{TB}{$\tan_\beta$}
\includegraphics[width=0.5\linewidth]{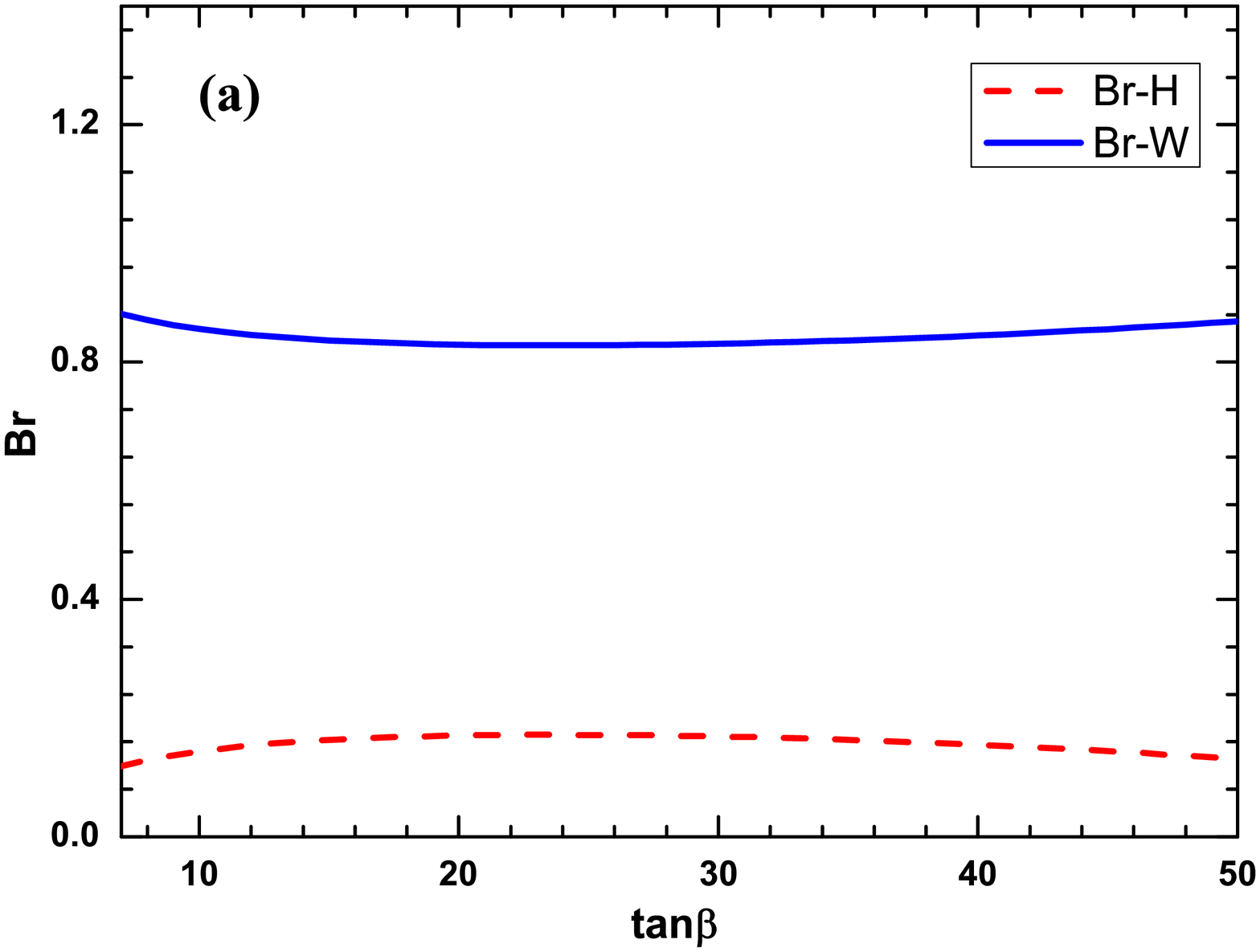}&
\includegraphics[width=0.5\linewidth]{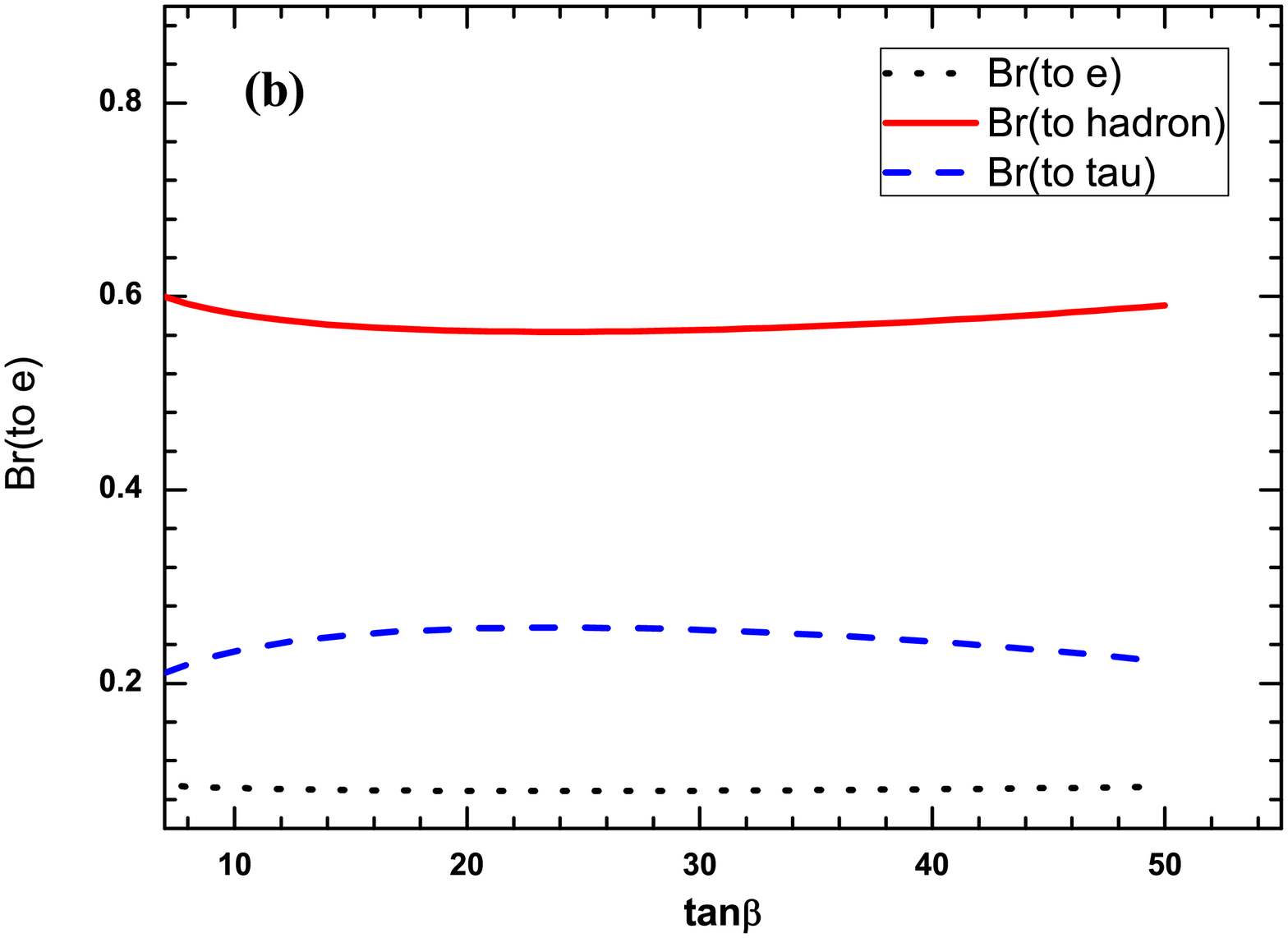}\\
\includegraphics[width=0.5\linewidth]{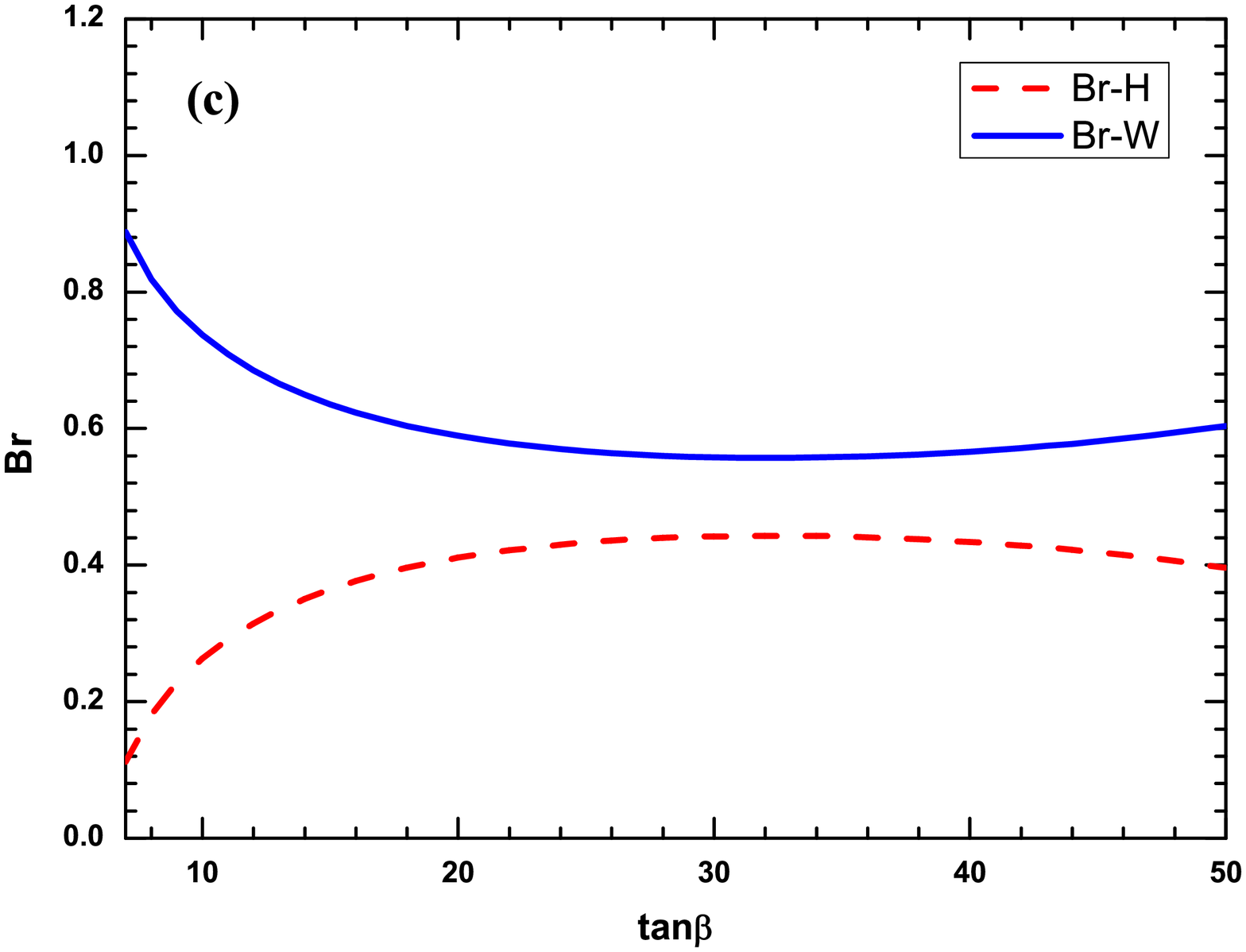}&
\includegraphics[width=0.5\linewidth]{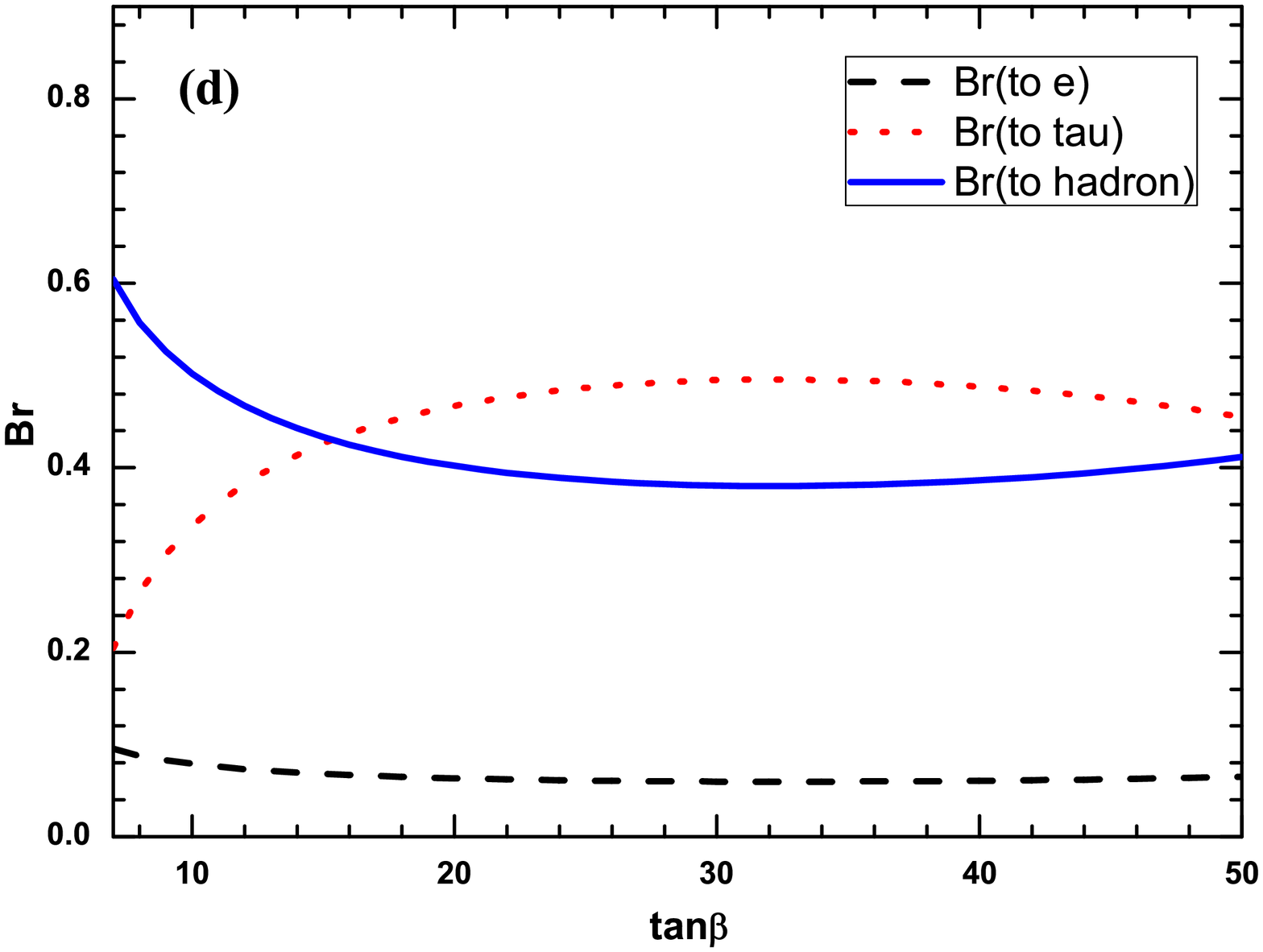}
\end{tabular}
\caption{ The $\tan\beta$-dependence of the branching ratios
$Br(\tilde \chi_1^- \to \tilde \chi_1^0 W^-)$, $ Br(\tilde \chi_1^- \to \tilde
\chi_1^0 H^-)$, $Br(\tilde \chi_1^- \to \tilde \chi_1^0 f \bar{f^\prime})$ with
 $f=e^-,\tau^-$ and hadrons, by assuming $(M_2,\mu)=(600,320)$ GeV ( the upper
 two figures (a) and (b)), or
 $(M_2,\mu)=(320,600)$ GeV ( the lower figures (c) and (d)).\label{fig:3}}
\end{figure}

Parameter $\tan\beta$ is one of the most important parameters in the SUSY models.
The $\tan\beta$-dependence of branching ratios of the lighter chargino $\tilde \chi_1^-$
decay modes is investigated here, as illustrated in Fig.\ref{fig:3},
where the branching ratios of $\tilde \chi_1^- \to \tilde
\chi_1^0 W^- (H^-)$ and $\tilde \chi_1^- \to \tilde \chi_1^0 f \bar{f^\prime}$
at one-loop level are shown with $7\leq \tan \beta \leq 50$.
We choose charged Higgs mass to be light so that $m_{H^-}< m_t + m_b$, therefore
the main decay mode of the charged Higgs is $\tau^-\bar\nu_\tau$, not $\bar t b$.

For the case of $(M_2,\mu)=(600,320)$ GeV, $\tilde \chi_1^- $ is higgsino-like, as illustrated by Fig.3(a) and 3(b), we find
 the following points:
\begin{itemize}
\item
The branching ratios of $\tilde \chi_1^-$ decays to $W^-$ and $H^-$  have a
rather weak dependence on $\tan \beta$, and
\beq
Br(\tilde \chi_1^- \to \tilde \chi_1^0 W^-)>
Br(\tilde \chi_1^- \to \tilde \chi_1^0 H^-),\label{eq:br1}
\eeq
in the whole region of $\tan\beta=[7,50]$.

\item
For the considered three body decays, there is the hierarchy
\beq
Br(\tilde \chi_1^- \to \tilde \chi_1^0 hadrons) >
Br(\tilde \chi_1^- \to \tilde \chi_1^0 \tau^- \bar\nu_\tau) >
Br(\tilde \chi_1^- \to \tilde \chi_1^0 l^- \bar\nu_l),\label{eq:br2}
\eeq
where $l=(e,\mu)$.
\end{itemize}

For the case of $(M_2,\mu)=(320,600)$ GeV, as illustrated by Fig.3(c) and 3(d), we find
a rather different picture from the case for  $(M_2,\mu)=(600,320)$ GeV:
For the region of large $\tan\beta$, say $\tan\beta > 15$, we find
\bea
Br(\tilde \chi_1^- \to \tilde \chi_1^0 W^-) & \gtrsim &
Br(\tilde \chi_1^- \to \tilde \chi_1^0 H^-),\label{eq:br3}\\
Br(\tilde \chi_1^- \to \tilde \chi_1^0 hadrons) & \lesssim &
Br(\tilde \chi_1^- \to \tilde \chi_1^0 \tau^- \bar\nu_\tau)
\gg Br(\tilde \chi_1^- \to \tilde \chi_1^0 l^- \bar \nu_l),\label{eq:br4}
\eea
with $l=(e,\mu)$. From Eqs.(\ref{eq:br1}-\ref{eq:br4}) one can see that the
pattern of the branching ratios  of the considered decays for two sets of input
parameters $(M_2,\mu)$ are rather different, which can be tested in the future
experiments.   Once people find branching ratios of $\tilde \chi_1^-$
decay modes are different from $W$ decays, one can believe that the charged Higgs
most possibly exist.

In Table \ref{table2}, we list the theoretical predictions for the branching ratios
of the three body decays $\tilde{\chi}_1^- \to  \tilde{\chi}_1^0 W^- (H^-)
\to  \tilde{\chi}_1^0 f f^\prime$ for the four sets of input parameters $(M_2,\mu)$,
and for $\tan\beta = 20$. Set-A: $(M_2,\mu)=(600,430)$ GeV;
Set-B: $(M_2,\mu)=(650,320)$ GeV; Set-C: $(M_2,\mu)=(320,650)$ GeV; and
Set-D: $(M_2,\mu)=(430,600)$ GeV.
\begin{table}[htb]
\begin{center}
\caption{Theoretical predictions for the branching ratios of the lighter chargino
three body decays $\tilde{\chi}_1^- \to
\tilde{\chi}_1^0 f f^\prime$ for the given values of $(M_2,\mu)$, and for $\tan\beta = 20$.
\label{table2} }
\vspace{0.3cm}
 \begin{tabular}{|c | c | c| c| c| c | c| c| c|}
 \hline
 	&	\multicolumn{4}{|c|}{Tree}	&	\multicolumn{4}{c|}{1-Loop}	\\
 \hline
 $\tilde{\chi}_1^- \to \tilde{\chi}_1^0ff^\prime$&	$\tilde{\chi}_1^0 e^-\bar{\nu}_e$& $\tilde{\chi}_1^0
 \mu^-\bar{\nu}_\mu$	&$\tilde{\chi}_1^0\tau^- \bar{\nu}_\tau$&	$\tilde{\chi}_1^0
 \; {\rm hadrons}$	&$\tilde{\chi}_1^0 \tilde{\chi}_1^0 e^-\bar{\nu}_e$
 & $\tilde{\chi}_1^0 \mu^-\bar{\nu}_\mu$	&	$\tilde{\chi}_1^0 \tau^-\bar{\nu}_\tau$	
 &	$\tilde{\chi}_1^0\; {\rm  hadrons}$	 \\
 \hline
 Set-A	&	0.0647	&	0.0661	&	0.4622	&	0.4070	&	0.0637	&	0.0640	&	0.4670	&	0.4063	\\
 \hline
Set-B	&	0.0875	&	0.0882	&	0.2895	&	0.5347	&	0.0875	&	0.0867	&	0.2738	&	0.5532	\\
 \hline
Set-C	&	0.0557	&	0.0574	&	0.5298	&	0.3570	&	0.0531	&	0.0540	&	0.5529	&	0.3409	\\
 \hline
Set-D  &	0.0381	&	0.0403	&	0.6632	&	0.2583	&	0.0359	&	0.0377	&	0.6916	&	0.2353	\\
 \hline
 \end{tabular}
\end{center}
\end{table}

In the framework of R-parity conserved MSSM, we study the higgsino and wino-like lighter
chargino decays to LSP and two SM fermions at one loop level.
The relevant SUSY parameters are chosen to make two body decay modes
$\tilde\chi_1^-\rightarrow \tilde \chi_1^0 W^- $ and
$\tilde\chi_1^-\rightarrow \tilde \chi_1^0 H^-$  are kinematically open
while others are closed. In this work the lightest neutralino is supposed to be
bino-like, and the charged Higgs boson mass is supposed to be lighter than
$m_t + m_b$. From our studies we find that
(a) the loop effects on the branching ratios are
small, less than $3\%$ in magnitude;
(b) the pattern of the decay rates of the considered decays on the choice of $(M_2,\mu)$
are specific and could be tested by future experiments, which could also be help for
searching for the signal of the charged Higgs boson.

For the light charged Higgs boson with mass lighter than the
top quark mass, it's dominant decay mode is $H^- \to \tau^- \nu_\tau$ with branching ratio
$\sim 100 \%$. The main background of charged Higgs production is $W$ boson,
which mainly decays to hadronic final states. Once branching ratio of
$\tilde \chi_1^-$ decays to $\tau^-\nu_\tau$ final state is found to be larger than
that of $W$ boson decays, it indicates that the charged Higgs boson may exist.
Suppose coupling of charged Higgs boson with $\tau^-\nu_\tau$ is well measured in other processes,
branching ratios of $\tilde \chi_1^-$ decays to $H^-$ and $W^-$, and hence the relation between
SUSY parameters can be well determined
from the branching ratio of $\tilde\chi_1^-$ decaying
to $\tilde\chi_1^0 \tau^- \nu_\tau$ final state.

\subsection*{Acknowledgments}
This work is supported by the National Natural Science Foundation of China
under Grants No. 11147023, 11305044 and 11235005,
and Zhejiang Provincial Natural Science Foundation of China under Grant No. LQ12A05003.


\end{document}